\documentclass[sn-mathphys-num,Numbered]{sn-jnl}

\usepackage{graphicx}%
\usepackage{multirow}%
\usepackage{amsmath,amssymb,amsfonts}%
\usepackage{amsthm}%
\usepackage{mathrsfs}%
\usepackage[title]{appendix}%
\usepackage{xcolor}%
\usepackage{textcomp}%
\usepackage{manyfoot}%
\usepackage{booktabs}%
\usepackage{algorithm}%
\usepackage{algorithmicx}%
\usepackage{algpseudocode}%
\usepackage{listings}%
\usepackage[symbol]{footmisc}
\usepackage{gensymb}

\usepackage[normalem]{ulem}

\theoremstyle{thmstyleone}%
\theoremstyle{thmstyletwo}%

\theoremstyle{thmstylethree}%

\begin{document}

\title[Demonstrating the ability of IceCube DeepCore to probe Earth's interior with atmospheric neutrino oscillations]{Demonstrating the ability of IceCube DeepCore to probe Earth's interior with atmospheric neutrino oscillations}

\author[1,a]{\fnm{Sharmistha} \sur{Chattopadhyay}}
\equalcont{(For the IceCube Collaboration\footnote[3]{\protect\url{http://icecube.wisc.edu} \\ $^\text{a}$also at Institute of Physics, Sachivalaya Marg, Sainik School Post, Bhubaneswar 751005, India.} ) }

\author* [1,a]{\fnm{Krishnamoorthi} \sur{J}}\email{kjayakumar@icecube.wisc.edu}
\equalcont{(For the IceCube Collaboration\footnote[3]{\protect\url{http://icecube.wisc.edu} \\ $^\text{a}$also at Institute of Physics, Sachivalaya Marg, Sainik School Post, Bhubaneswar 751005, India.} ) }

\author[1,a]{\fnm{Anuj Kumar} \sur{Upadhyay}}

\equalcont{(For the IceCube Collaboration\footnote[3]{\protect\url{http://icecube.wisc.edu} \\ $^\text{a}$also at Institute of Physics, Sachivalaya Marg, Sainik School Post, Bhubaneswar 751005, India.} ) }

\affil[1]{\orgdiv{Dept. of Physics and Wisconsin IceCube Particle Astrophysics Center}, \orgname{University of Wisconsin-Madison}, \orgaddress{\city{Madison, WI 53706},  \country{USA}}}

\abstract{The IceCube Neutrino Observatory is an optical Cherenkov detector instrumenting one cubic kilometer of ice at the South Pole. The Cherenkov photons emitted following a neutrino interaction are detected by digital optical modules deployed along vertical strings within the ice. The densely instrumented bottom central region of the IceCube detector, known as DeepCore, is optimized to detect GeV-scale atmospheric neutrinos. As upward-going atmospheric neutrinos pass through Earth, matter effects alter their oscillation probabilities due to coherent forward scattering with ambient electrons. These matter effects depend upon the energy of neutrinos and the density distribution of electrons they encounter during their propagation. Using simulated data at the IceCube Deepcore equivalent to its 9.3 years of observation, we demonstrate that atmospheric neutrinos can be used to probe the broad features of the Preliminary Reference Earth Model. In this contribution, we present the preliminary sensitivities for establishing the Earth matter effects, validating the non-homogeneous distribution of Earth’s electron density, and measuring the mass of Earth. Further, we also show the DeepCore sensitivity to perform the correlated density measurement of different layers incorporating constraints on Earth's mass and moment of inertia.}


\maketitle

\section{Introduction}\label{sec:intro}
The phenomenon of neutrino flavor oscillations,  described by oscillation parameters such as mixing angles ($\theta_{12}$, $\theta_{13}$, and $\theta_{23}$) and mass-squared splittings ($\Delta m^2_{21}$ and $\Delta m^2_{31}$), has been observed by several experiments involving solar, atmospheric, reactor, and accelerator neutrinos. These observations suggest that neutrinos must have non-zero masses and mix with each other. The neutrino mixing can be explained by representing neutrino flavor eigenstates ($\nu_e$, $\nu_\mu$, $\nu_\tau$) as a superposition of their mass eigenstates ($\nu_1$, $\nu_2$, $\nu_3$) through three mixing angles ($\theta_{12}$, $\theta_{13}$, and $\theta_{23}$), and one CP-violating Dirac phase $\delta_\text{CP}$. The standard parametrization of these mixing parameters is described by a unitary matrix known as the Pontecorvo-Maki-Nakagawa-Sakata (PMNS) matrix.
Using data from various neutrino oscillation experiments, $\theta_{12}$, $\theta_{13}$, $\Delta m^2_{21}$, and $|\Delta m^2_{31}|$ have been measured with a precision of a few percentages, whereas there are still three unknowns that need to be resolved, namely, the value of $\delta_\text{CP}$,  the octant of $\theta_{23}$, and the neutrino mass ordering~\cite{Esteban:2024eli}. The upcoming neutrino experiments can potentially determine the oscillation parameters and resolve these three unknowns with great precision. The precise measurement of oscillation parameters, particularly the non-zero value of $\theta_{13}$ discovered by the Daya Bay experiment in 2012~\cite{DayaBay:2012fng}, has opened a new era of determining the neutrino mass ordering, probing several interesting beyond the Standard Model scenarios, and unravelling the internal structure of deep Earth through the matter effects experienced by the upward-going atmospheric neutrinos as they travel through Earth.

The internal structure of Earth has traditionally been studied using indirect methods such as gravitational~\cite{Chen:2014,astro_almanac} and seismic studies~\cite{McDonough_MMTE:2023,McDonough:2024}. The seismic studies indicate that Earth has a layered structure that can be broadly classified into two concentric shells - a high-density core and a low-density mantle. Based on the seismic wave propagation data, the widely accepted density model of Earth, known as the Preliminary Reference Earth Model (PREM)~\cite{Dziewonski:1981xy}, has been developed. However, independent and complementary information can be obtained using the weakly interacting neutrinos as they pass through Earth. As multi-GeV neutrinos propagate through Earth, they undergo coherent forward scattering with the ambient electrons, which alters their oscillation probabilities. These are known as the matter effects, which depend upon the density distribution of electrons along the neutrino path. Therefore, matter effects in neutrino oscillations can be used to probe the internal structure of Earth. Recently, several sensitivity studies have been performed to probe the interior of Earth by utilizing the matter effects in neutrino oscillations at various upcoming atmospheric neutrino experiments such as ORCA~\cite{Bourret:2017tkw,Capozzi:2021hkl,Maderer:2022toi,DOlivoSaez:2022vdl}, DUNE~\cite{Kelly:2021jfs,Denton:2021rgt}, ICAL~\cite{Kumar:2021faw,Upadhyay:2021kzf,Upadhyay:2022jfd,Raikwal:2023jkf,Upadhyay:2024gra}, and Hyper-K~\cite{Rott:2015kwa,Jesus-Valls:2024tgd}. The neutrino interaction cross section increases with the energy of neutrinos, and at energies above a few TeV, the cross section is high enough to cause a noticeable attenuation in the neutrino flux as it passes through Earth~\cite{Gandhi:1995tf,IceCube:2017roe}. The initial studies of exploiting attenuation of high energy neutrino flux to explore the Earth's interior have been discussed in Refs.~\cite{DeRujula:1983ya,Wilson:1983an,Askarian:1984xrv,Volkova:1985zc,Tsarev:1985yub,Borisov:1986sm,Tsarev:1986xg,Jain:1999kp,Winter:2006vg,Kuo:1995,Crawford:1995,Reynoso:2004dt}. In Ref.~\cite{Gonzalez-Garcia:2007wfs}, the authors forecasted that IceCube could reject the hypothesis of Earth's homogeneity at a $3.4\sigma$ confidence level using ten years of high-energy atmospheric neutrino absorption data. Additionally, Ref.~\cite{Donini:2018tsg} utilized one year of IceCube's multi-TeV atmospheric muon neutrino data to estimate the densities of various Earth layers and measured the mass and the moment of inertia of Earth via absorption of high-energy neutrinos. 

In this work, we use the atmospheric neutrino data in the multi-GeV energy range simulated for the IceCube DeepCore detector equivalent to 9.3 years of observation. Using this low-energy simulated data, we aim to establish IceCube DeepCore's sensitivity to Earth matter effects in atmospheric neutrino oscillations. We demonstrate, the ability of the IceCube DeepCore data to validate the non-uniform distribution of matter inside Earth and measure the total amount of matter or mass of Earth. We further evaluate the sensitivity to measure the density of different layers inside Earth while assuming constraints from Earth's mass and moment of inertia. 

\begin{figure}
    \centering
    \includegraphics[width=0.49\linewidth]{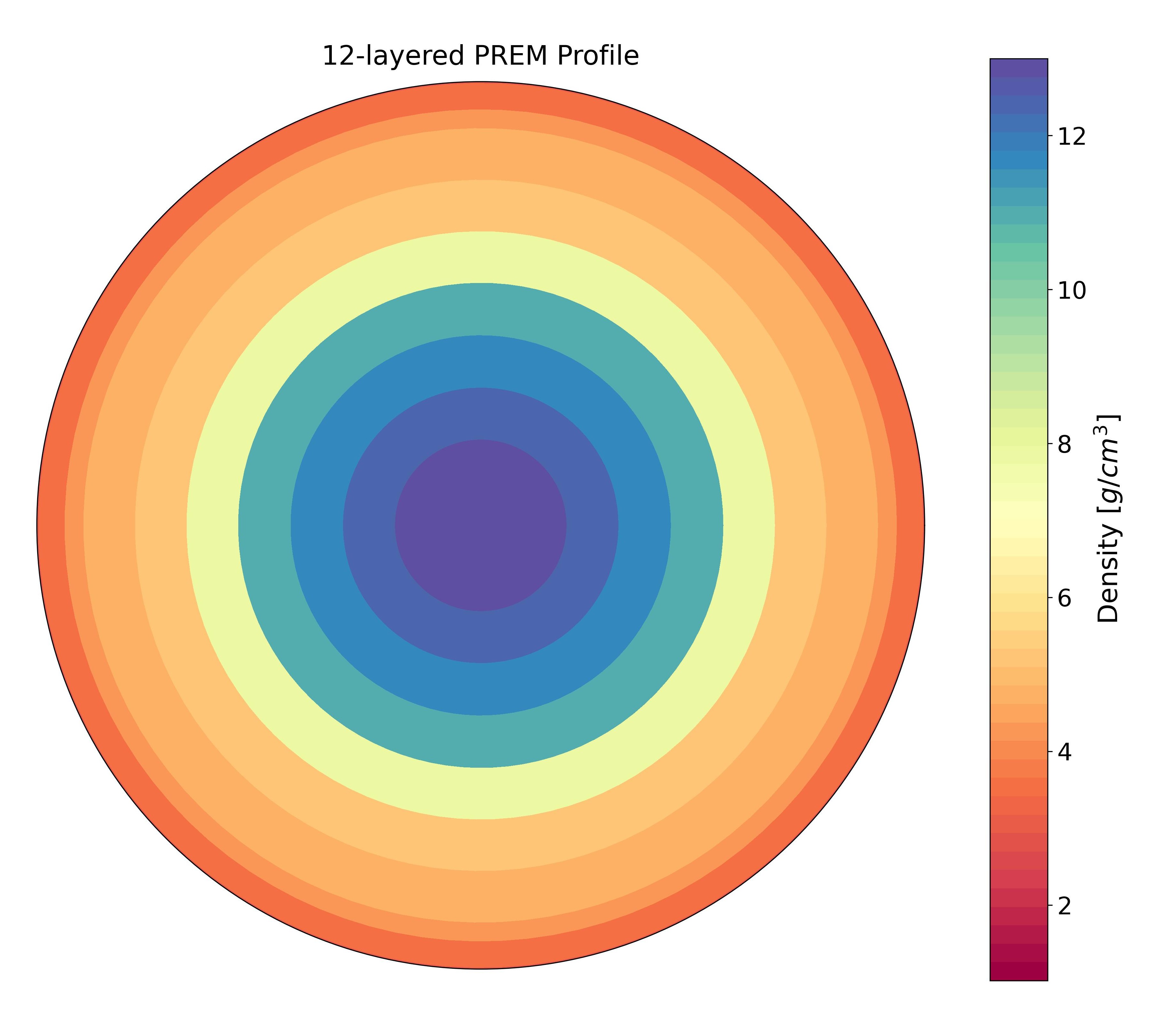}
    \includegraphics[width=0.49\linewidth]{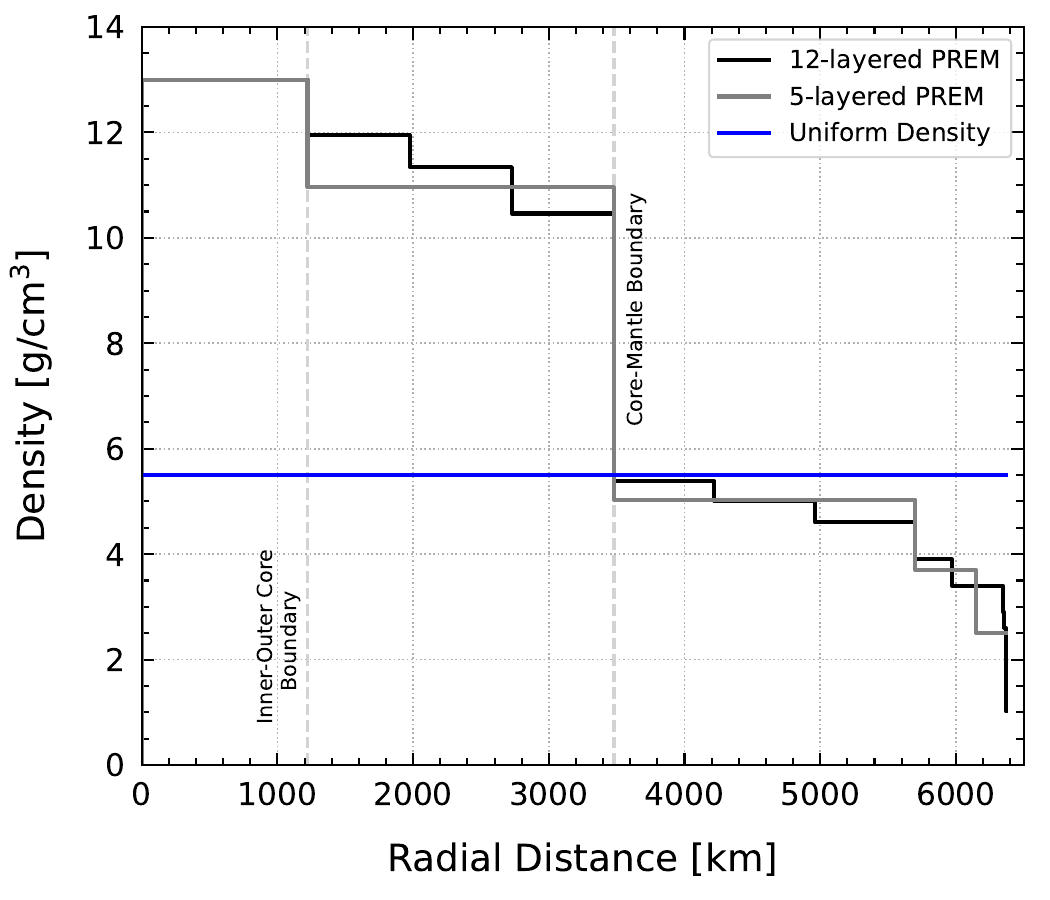}
    \caption{Left: Radial distribution of Earth's density using the 12-layered PREM model. Right: Various profiles of radial density distribution inside Earth such as 12-layered and 5-layered PREM profiles, and a uniform density profile.}
    \label{fig:density_profiles}
\end{figure}

We perform three distinct matter-effect-related analyses, referred to as Analyses I, II, and III. We evaluate a specific Earth density model by comparing it to the 12-layered (5-layered for correlated density measurement) matter density model guided by the PREM profile. The left panel of Figure~\ref{fig:density_profiles} shows the density model of Earth with 12 layers, where density within a layer remains constant, while the right panel of Fig.~\ref{fig:density_profiles} depicts the density distribution as a function of radial distance from Earth's center for the considered density models. The black, grey, and blue curves correspond to the 12-layered, 5-layered, and uniform density profiles of Earth, respectively. Now, we brieflly discuss about the three analyses that we present in this proceeding.

\medskip
\textit{\textbf{Analysis I:}} In this analysis, we aim to reject a vacuum oscillation scenario with respect to oscillations in the presence of matter as governed by the 12-layered PREM profile of Earth.

\medskip
\textit{\textbf{Analysis II:}} This analysis aims to reject a uniform density profile with respect to a non-uniform (12-layered PREM) density profile.

\medskip
\textit{\textbf{Analysis III:}} This analysis has two parts. In the first part, we aim to measure the mass of Earth by considering a 12-layered density profile. The density of each layer is scaled by a single scaling factor ($\alpha$), which is varied without applying external constraints from seismic and gravitational studies.

In the second part of the analysis, we measure the correlated density of Earth's layers considering a simplified 5-layered density profile, where the five layers are inner core (IC), outer core (OC), inner mantle (IM), middle mantle (MM), and outer mantle (OM). We consider the following constraints in this analysis: (i) the mass and moment of inertia of Earth are fixed, (ii) the density of OM is fixed, and (iii) the ratio of densities between the IC and OC has been taken to be the same as their ratio in the PREM profile. After applying these constraints, the variations in the densities of the IC, OC, IM, and MM layers can be parameterized by a single scaling factor $(\alpha_c)$. We treat the scaling factors ($\alpha$ and $\alpha_c$) as observables in the respective parts of the analysis, which we aim to measure using the IceCube DeepCore oscillation data.

In these analyses, we have utilized simulated IceCube DeepCore data to develop a robust analysis setup, which will be used to perform the analyses with the real experimental data.

\section{Earth matter effects in atmospheric neutrino oscillations}
\label{sec:matter_effects}

Atmospheric neutrinos are produced by the interactions of cosmic rays in Earth's atmosphere, spanning energy ranges from a few MeV to hundreds of TeV and travelling across wide baselines, from 20 km to 13,000 km. The flux consists of electron and muon neutrinos and their antineutrinos, which undergo oscillations at multi-GeV energies. As upward-going multi-GeV neutrinos pass through Earth, they experience charged-current (CC) interactions due to coherent forward scattering with ambient electrons. The effective CC matter potential for $\nu_e (+)$ and $\Bar{\nu}_e (-)$ is given by:
\begin{align}
\begin{split}
V_\text{CC} &= \pm\, \sqrt{2} G_F N_e  \approx \pm \, 7.6 \times Y_e \times 10^{-14} \left[\frac{\rho}{\text{g/cm}^3}\right]~\text{eV}\,,
\end{split}
\end{align}
where $G_F$ is the Fermi coupling constant, $N_e$ is electron number density, and $Y_e = N_e/(N_p + N_n)$ is electron-to-nucleon fraction inside matter with density $\rho$. Here, $N_p$ and $N_n$ represent the proton and neutron number densities, respectively. The following $Y_e$ values for the PREM profile are considered: 0.4656 for both the inner and outer core and 0.4957 for the mantle as given in Ref.~\cite{Rott:2015kwa}. For the uniform matter density profile, $Y_e$ is considered to be 0.5, assuming Earth is neutral and isoscalar. This matter potential modifies the effective neutrino mass-splittings and mixing angles, thereby altering the oscillation probabilities as neutrinos propagate through Earth.

\begin{figure}
    \centering
    \includegraphics[width=1\linewidth]{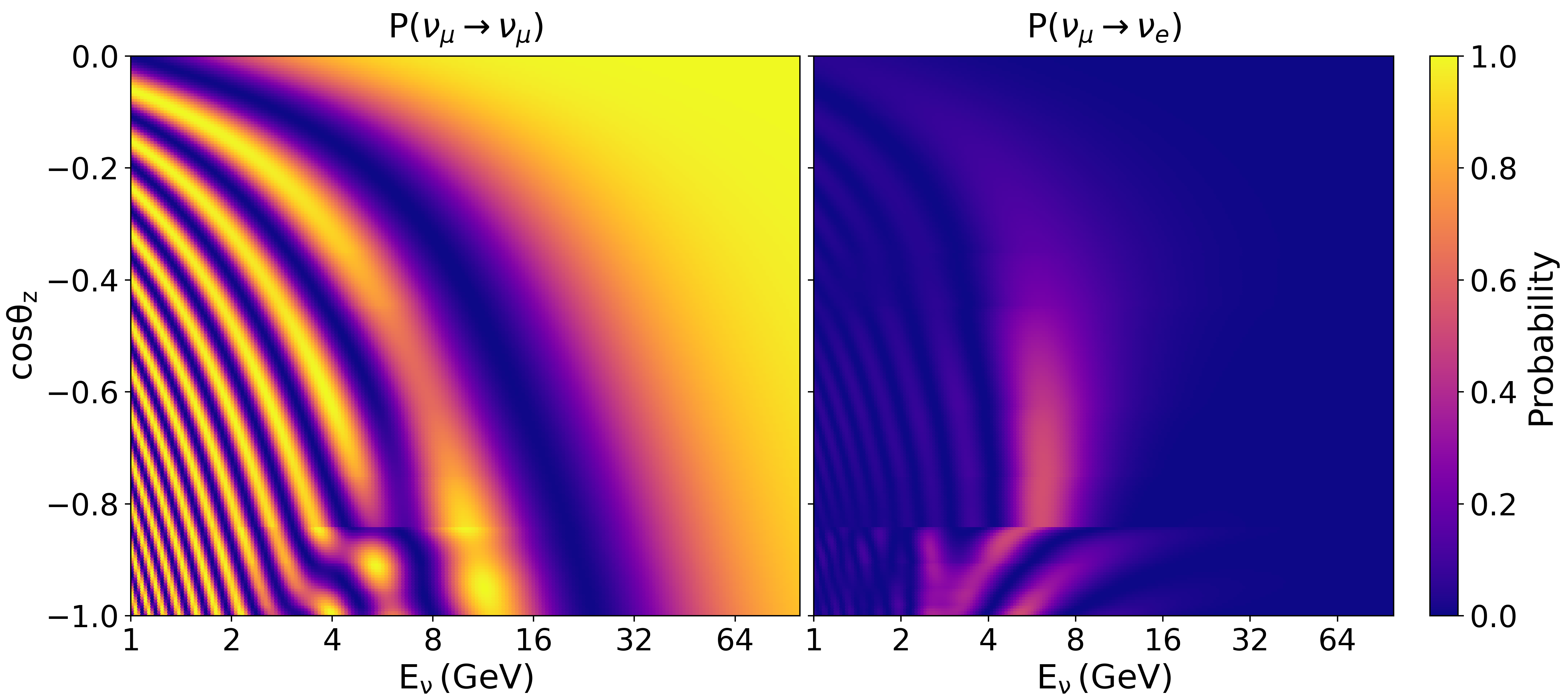}
    \caption{Three-flavor $P(\nu_\mu \rightarrow \nu_\mu)$ survival probabilities (left panel) and $P(\nu_\mu \rightarrow \nu_e)$ appearance probabilities (right panel) in the $(E_\nu, \cos\theta_\nu)$ plane, calculated using the PREM density model of Earth and the values of neutrino oscillation parameters consistent with the global fit assuming normal mass ordering ($m_3\,>\,m_2\,>\,m_1$). We assume $\theta_{13}=8.54\degree$, $\theta_{23}=47.5\degree$, and $\Delta m_{31}^2 = 2.47\times10^{-3}$ eV$^2$ to prepare this plot.}
    \label{fig:PREM_osc}
\end{figure}

The resonant enhancement in the effective value of the smallest neutrino mixing angle, $\theta_{13}$, which can increase to as much as $45^\circ$ due to the Earth matter effects, is known as the Mikheyev-Smirnov-Wolfenstein (MSW) resonance~\cite{Wolfenstein:1977ue,Mikheev:1986gs,Mikheev:1986wj}. This resonance is prominent for neutrinos in the case of normal mass ordering (NO) and for antineutrinos in the case of inverted mass ordering (IO). The MSW resonance depends upon the energy of the neutrino and the electron density it encounters along its path. Furthermore, neutrinos that pass through the Earth's core encounter a sharp density transition at the core-mantle boundary, causing significant changes in the oscillation probabilities. This phenomenon is referred to as the parametric resonance (PR)~\cite{Ermilova:1986,Akhmedov:1988kd,Krastev:1989,Akhmedov:1998ui,Akhmedov:1998xq} or neutrino oscillation length resonance (NOLR)~\cite{Petcov:1998su,Chizhov:1998ug,Petcov:1998sg,Chizhov:1999az,Chizhov:1999he}. These matter resonances are shown in Fig.~\ref{fig:PREM_osc}, which presents the $P(\nu_\mu \rightarrow \nu_\mu)$ survival probability and $P(\nu_\mu \rightarrow \nu_e)$ appearance probability in the $(E_\nu, \cos\theta_\nu)$ plane using three-flavor neutrino oscillations in the presence of Earth matter effects by considering the PREM density model of Earth.

\section{Reconstruction and event selection at IceCube DeepCore}
\label{sec:reconstruction}

\begin{figure}
	\centering
	\includegraphics[width=1\linewidth]{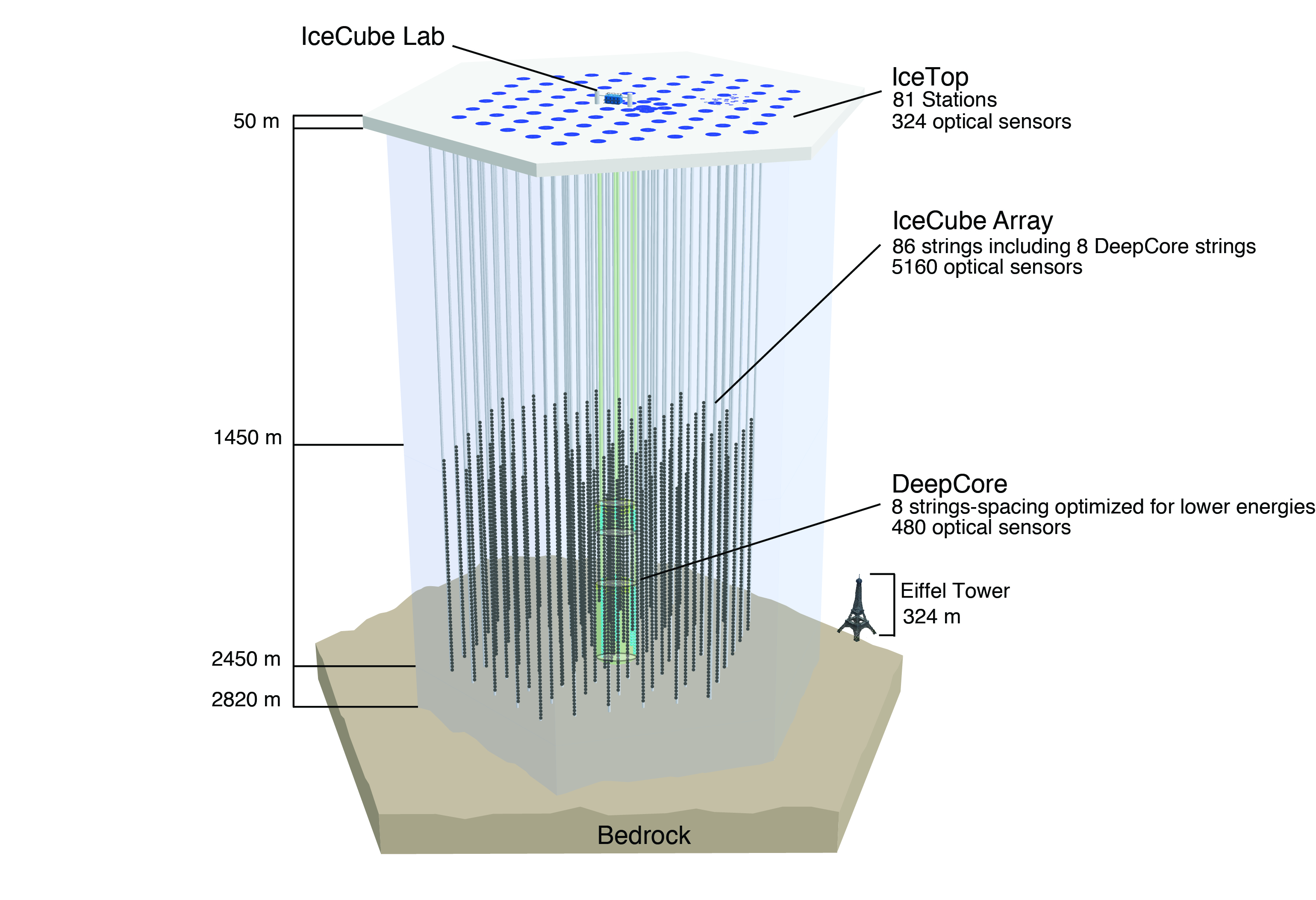}
	\caption{Schematic image of the IceCube Neutrino Observatory, which includes the main IceCube detector, a denser array called DeepCore for low-energy neutrino detection, and a surface detector array known as IceTop for atmospheric muon veto.}
	\label{fig:detector_layout}
\end{figure}

The IceCube Neutrino Observatory~\cite{IceCube:2016zyt} is located 1.5 km beneath the South Pole ice, covering a cubic kilometer of ice with 5160 Digital Optical Modules (DOMs) deployed along 86 vertical strings, as illustrated in Fig.~\ref{fig:detector_layout}. These DOMs are highly sensitive to Cherenkov photons emitted by charged particles resulting from neutrino interactions in the ice. The densely instrumented bottom central region of the IceCube detector, the DeepCore sub-array~\cite{IceCube:2011ucd}, comprises 15 densely packed strings, consisting of 7 IceCube strings and 8 DeepCore strings, each separated by approximately 70 meters, with a vertical spacing of 7 meters between each DOM. This sub-array enhances the detector's capability to identify neutrinos at the GeV scale, where the Earth matter effects on neutrino oscillations become significant.

The digitized signal from the DOMs contains information on the charge and the arrival time of photons emitted by neutrino-induced charged particles. This information is used to reconstruct the properties of interacting neutrinos, such as their energy and direction of arrival. The collected data is processed through multiple filter levels to reduce the background, which consists of atmospheric muons and random detector noise. There are a total of five levels of filtering explained in detail in Ref.~\cite{IceCubeCollaboration:2023wtb}. 

After applying these filters, the final sample has a significantly reduced background, less than 1\% of the sample, making it suitable for applying the reconstruction algorithm. A Convolutional Neural Networks (CNNs) based machine learning algorithm was utilized to reconstruct neutrino observables~\cite{IceCube:2024xjj}. Five variables are used as input to the CNNs: (i) total charge, (ii) time of the first hit, (iii) time of the last hit, (iv) time-weighted mean of charges, and (v) time-weighted standard deviations of charges. The CNNs reconstruct various neutrino properties, including neutrino energy, arrival angle with respect to the zenith ($\theta_\text{zenith}$), interaction vertex position (x, y, z), particle identification (PID), and a classifier for rejecting atmospheric muon backgrounds. A PID classifier is used to identify track-like topology formed by $\nu_{\mu}$ CC interactions, and cascade-like topology arising from $\nu_{e}$ CC,  and all neutral-current (NC) interactions. Further, DeepCore is also sensitive to $\nu_{\tau}$ CC interactions which contributes to track-like as well as cascade-like topologies. The PID classifier assigns a score between 0 and 1 for each event. A score close to 1 indicates a high probability that the observed event is a $\nu_{\mu}$ CC event. 

\begin{figure}
    \centering
    \includegraphics[width=1\linewidth]{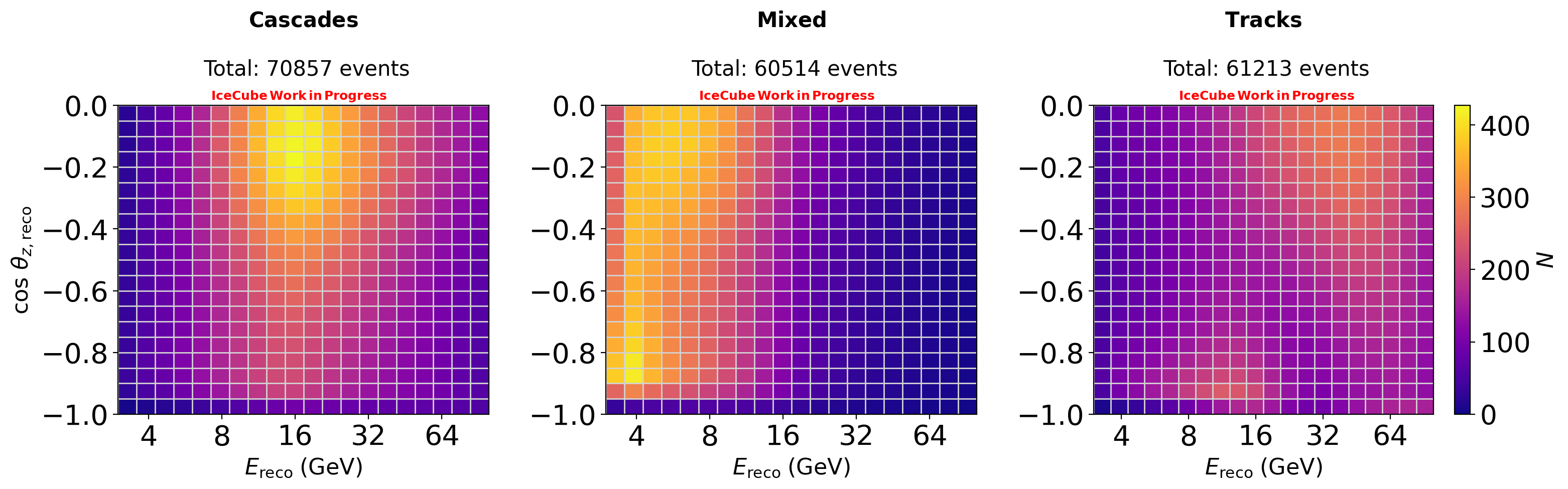}
    \caption{The expected number of events based on the 12-layered PREM profile of Earth, as a function of reconstructed energy and $\cos\theta_\text{zenith}$ of neutrino for each PID in 9.3 years of lifetime. Each histogram represents one PID bin, defined by the PID boundaries. The boundaries of PID bins are 0 to 0.33 for cascades, 0.33 to 0.39 for mixed, and 0.39 to 1 for tracks.}
    \label{fig:event_distribution}
\end{figure}

Additionally, the following final level cuts are applied to select the analysis events; the reconstructed neutrino interaction vertex is within DeepCore, the reconstructed energy is between 3 to 100 GeV, and the reconstructed $\cos\theta_\text{zenith}$ is below 0.
The selected sample is binned using 20 logarithmically-spaced bins of reconstructed energy from 3 to 100 GeV, 20 linear-spaced bins of reconstructed $\cos\theta_\text{zenith}$ between [-1, 0], and 3 PID bins with bin edges of [0, 0.33, 0.39, 1] corresponding to cascade-, mixed-, and track-like topologies. The distribution of expected events as a function of reconstructed energy and $\cos\theta_\text{zenith}$ of neutrino for each PID is shown in Fig.~\ref{fig:event_distribution}. In 9.3 years, with this choice of analysis range, the expected neutrino events are around 192 k. For Analysis I and Analysis III, energy bins below 5 GeV are not considered to make these analyses more robust. 

\section{Numerical analysis}
\label{sec:analysis}

In the following, we describe the analysis methodology to perform the sensitivity studies for the Earth matter effects-related analyses using simulated Monte Carlo (MC) events without any statistical fluctuations, also known as `Asimov dataset'~\cite{Cowan:2010js}. In order to perform these analyses, the MC events are re-weighted with oscillation probabilities, which are calculated assuming different Earth density models. In these analyses, we also incorporate systematic uncertainties in terms of nuisance parameters, which include uncertainties in atmospheric neutrino flux, neutrino oscillation parameters, cross section, and detector response. These nuisance parameters are described in detail in Refs.~\cite{IceCubeCollaboration:2023wtb,IceCube:2024xjj}.

A fit is performed over nuisance parameters using a binned log-likelihood ($LLH$) which is defined as:
\begin{equation}
LLH = -\sum_{i\in bins} \log \Big(\frac{n_{i}^{ n_{o}} e^{-n_{i}}}{n_{o}!}\Big) + \sum_{j \in \mathrm{syst}}^{}\frac{(\hat{p}_j - {p_j})^2}{2\sigma^2_{j}}\,,
\label{eq:metric}
\end{equation}
where the first term represents the negative of the Poisson log-likelihood. 
Here, $n_i$ ($n_o$) is the number of expected (observed) events in the $i^{th}$  bin. The second term is the sum over pull penalties where the pull penalty for the $j^{th}$ nuisance parameter $(p_j)$ is defined in terms of the Gaussian prior $\sigma_{j}$ around its central value $\hat{p}_j$.

In general, the hypothesis test is commonly performed by following the Wilks' theorem~\cite{wilks1938}. However, rejecting a given (vacuum or uniform) density profile with respect to the PREM profile is a binary hypothesis test. Therefore, the Wilks' theorem cannot be applied as they are not connected by a continuous parameter. The sensitivity for non-nested hypotheses is estimated following Ref.~\cite{Ciuffoli:2013rza} and given by:
\begin{equation}
\eta_{\sigma} = \frac{\Delta LLH_\text{true} + \Delta LLH_\text{alternative}}{\sqrt{2\times\Delta LLH_\text{alternative}}} \,,
\label{eq:sen_non_nested}
\end{equation}
where \(\Delta LLH_\text{true}\) is the difference in the log-likelihood values when the alternative hypothesis is compared with the true hypothesis. Mathematically, it is expressed as:
\begin{equation}
\Delta LLH_\text{true} = LLH_\text{true}^\text{alternative} - LLH_\text{true}^\text{true}\,,
\end{equation}
where \(LLH_\text{true}^\text{true}\) represents the log-likelihood value obtained when the MC data is generated using the true hypothesis and maximized using the true hypothesis. The \(LLH_\text{true}^\text{alternative}\) represents the log-likelihood value when the same MC data, generated using the true hypothesis, is maximized using the alternative hypothesis. Similarly, the \(\Delta LLH_\text{alternative}\) is the log-likelihood difference when the MC data is generated using the alternative hypothesis with the best-fit parameters obtained from \(LLH_\text{true}^\text{alternative}\), and maximized using the true and the alternative hypothesis. It is expressed as:
\begin{equation}
\Delta LLH_\text{alternative} = LLH_\text{true}^\text{alternative} - LLH_\text{alternative}^\text{alternative} \,.
\end{equation}

On the other hand, the determination of Earth's mass can be derived by varying a continuous parameter (denoted by \(\alpha\)), making it a case of nested hypothesis testing, and the corresponding sensitivity is calculated following Wilks' theorem and given by:
\begin{equation} 
\eta^m_{\sigma} = \sqrt{2 \times \Delta LLH} \,, 
\label{eq:sens_nested} 
\end{equation} 
where \(\Delta LLH = LLH(\alpha) - LLH(\alpha_0)\). Here, the MC data is generated using the 12-layered PREM hypothesis and maximized with respect to the hypotheses with scaling factors $\alpha$ and $\alpha_0$, which correspond to the modified and standard Earth models, respectively.

Similarly, for the correlated density measurement case, sensitivity is again obtained through a nested hypothesis testing approach, with a continuous parameter denoted by \(\alpha^c\). The corresponding sensitivity is defined by: 
\begin{equation} 
\eta^c_{\sigma} = \sqrt{2 \times \Delta LLH^c} \,, 
\label{eq:sens_nested2} 
\end{equation}
where, \(\Delta LLH^c = LLH(\alpha^c) - LLH(\alpha^c_0)\). In this scenario, MC data is generated based on a 5-layered PREM model. Hypothesis testing then involves maximizing likelihoods with respect to scaling factors \(\alpha^c\) for the modified Earth model and \(\alpha^c_0\) for the standard 5-layered model.
As the sensitivities in Eq.~\ref{eq:sen_non_nested}, Eq.~\ref{eq:sens_nested}, and Eq.~\ref{eq:sens_nested2} are calculated using the Asimov dataset, we are referring to them as `Asimov sensitivity' throughout the paper.

\section{Sensitivity results}
\label{sec:results}

This section presents the Asimov sensitivities for all three matter-effects-related analyses. Currently, $\theta_{23}$ is the most uncertain parameter apart from $\delta_\text{CP}$, and the sensitivity to matter effects depends on the true choice of $\theta_{23}$. Therefore, we present the sensitivity for Analysis I and Analysis II as a function of $\sin^2\theta_{23}$ (true). During the fit, $\Delta m^2_{31}$ and $\theta_{23}$ have been kept free along with other nuisance parameters, while the remaining oscillation parameters; $\theta_{12}$, $\Delta m^2_{21}$, and $\theta_{13}$ are kept fixed, as they have been precisely measured. We also assume that $\delta_\text{CP} = 0$ and keep it fixed, as its impact on these analyses is minimal.

\subsection{Establishing Earth matter effects}
\label{sec:analysis_I}

\begin{figure}
	\centering
	\includegraphics[width=0.495\linewidth]{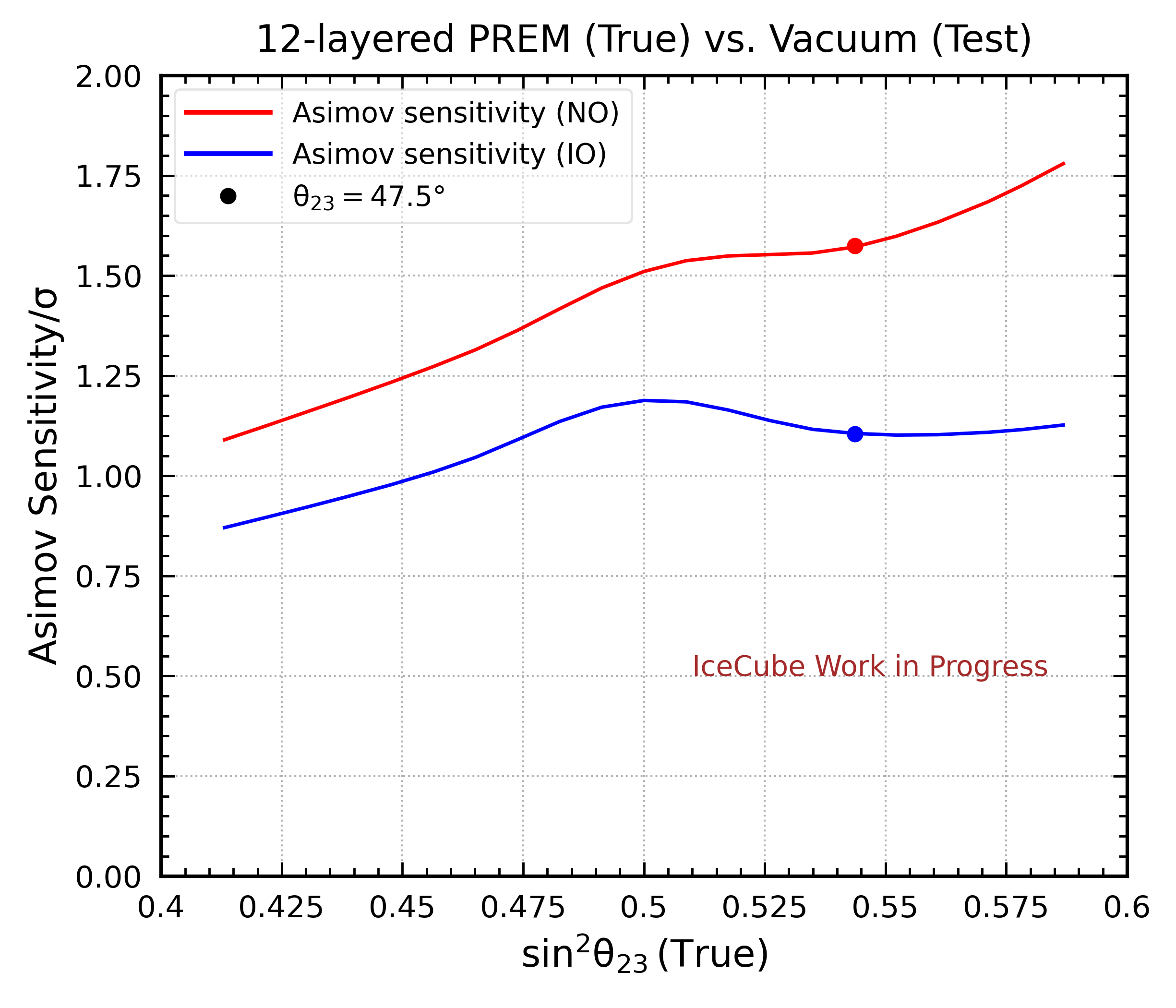}
        \includegraphics[width=0.495\linewidth]{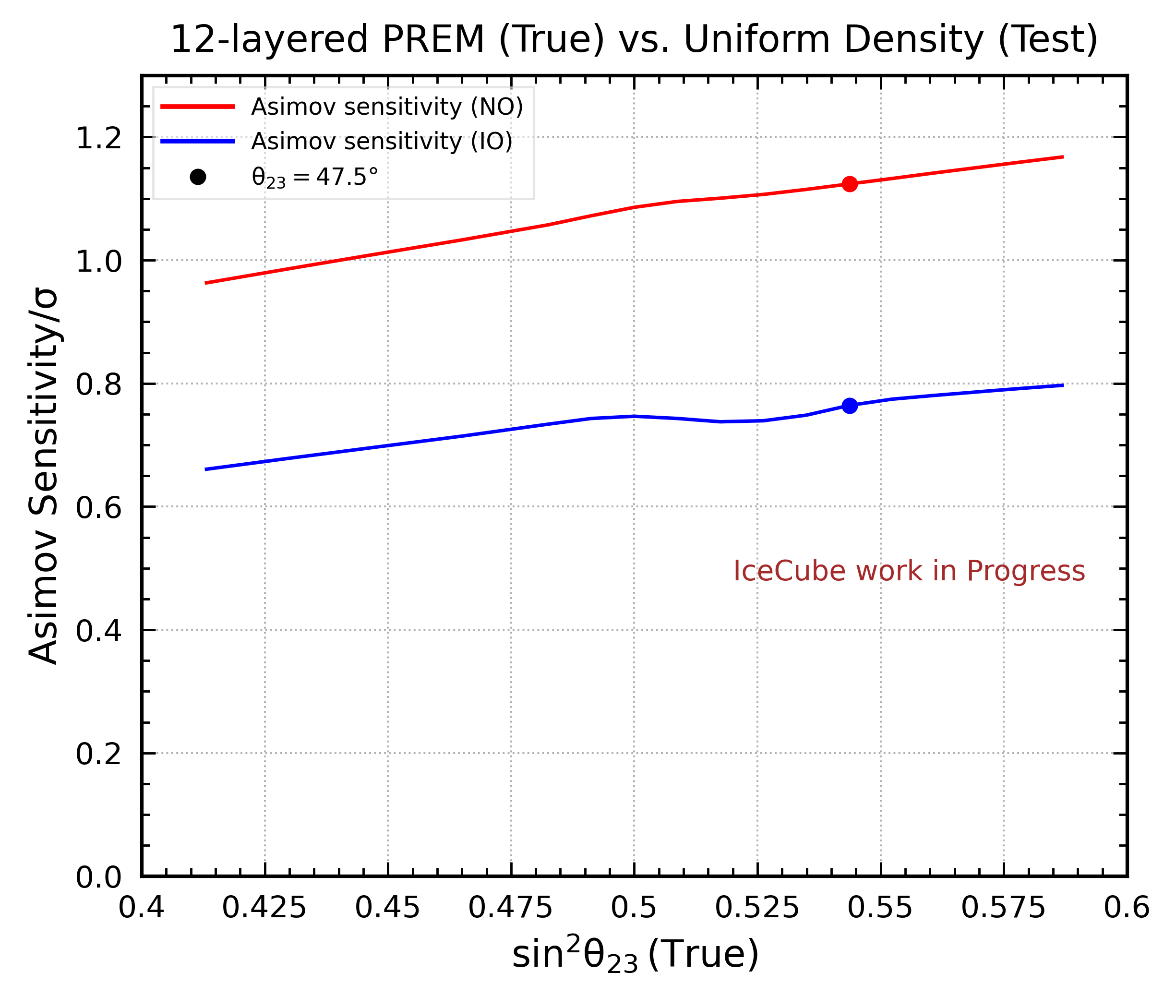}
	\caption{Left: Asimov sensitivity as a function of simulated $\sin^2\theta_{23}$ (true) to establish Earth matter effects in atmospheric neutrino oscillations by rejecting the vacuum oscillations with respect to oscillations in the presence of Earth's matter. Right: Asimov sensitivity as a function of simulated $\sin^2\theta_{23}$ (true) to validate the layered structure inside Earth by rejecting the uniform matter density profile with respect to the PREM density profile.}
	\label{fig:results_analysis1_analysis2}
\end{figure}

In this section, we present the Asimov sensitivity for establishing Earth matter effects in atmospheric neutrino oscillations by rejecting the vacuum hypothesis with respect to the PREM hypothesis. The left panel of Fig.~\ref{fig:results_analysis1_analysis2} presents the sensitivity as a function of $\sin^2\theta_{23}$ (true). The red (blue) curve represents the sensitivity assuming normal (inverted) neutrino mass ordering in both simulation (true) and analysis (test). The linear behaviour of the sensitivity as a function of $\sin^2\theta_{23}$ is due to the dominant contribution of matter effects in the $P(\nu_\mu \rightarrow \nu_\mu)$ survival probability and the $P(\nu_\mu \rightarrow \nu_e)$ appearance probability being proportional to $\sin^2\theta_{23}$, as shown by series expansion in Ref.~\cite{Akhmedov:2004ny}. The red (blue) dot indicates the sensitivity for a representative true choice of $\theta_{23} = 47.50^\circ$ corresponding to $1.57 \sigma$ for NO ($1.1 \sigma$ for IO).

\subsection{Validating layered structure inside Earth}

In this section, we present the significance of the IceCube DeepCore detector in validating the layered structure inside Earth by rejecting the uniform matter density profile with respect to the PREM density profile. The right panel of Fig.~\ref{fig:results_analysis1_analysis2} presents the Asimov sensitivity as a function of the simulated $\sin^2\theta_{23}$ (true). The red (blue) curve represents the sensitivity, assuming normal (inverted) mass ordering in both simulation (true) and analysis (test). The linear behaviour of the sensitivity as a function of $\sin^2\theta_{23}$ is due to the same reason as explained in section~\ref{sec:analysis_I}. The sensitivity for a representative true choice of $\theta_{23} = 47.50^\circ$ assuming normal (inverted) mass ordering is $1.12 \sigma$ ($0.76 \sigma$).

\subsection{Measuring the mass of Earth}
In this section, we present the Asimov sensitivity for the measurement of the mass of Earth. The left panel of Fig.~\ref{fig:results_analysis3} illustrates the sensitivity to Earth's mass using a 12-layered PREM profile, without applying any external constraints. This approach does not seek to improve existing gravitational measurements, but rather tries to demonstrate how effectively neutrinos, through weak interactions alone, can independently estimate Earth's mass.

\subsection{Correlated density measurement of various layers inside Earth}
\begin{figure}
\centering
\includegraphics[width=0.455\linewidth]{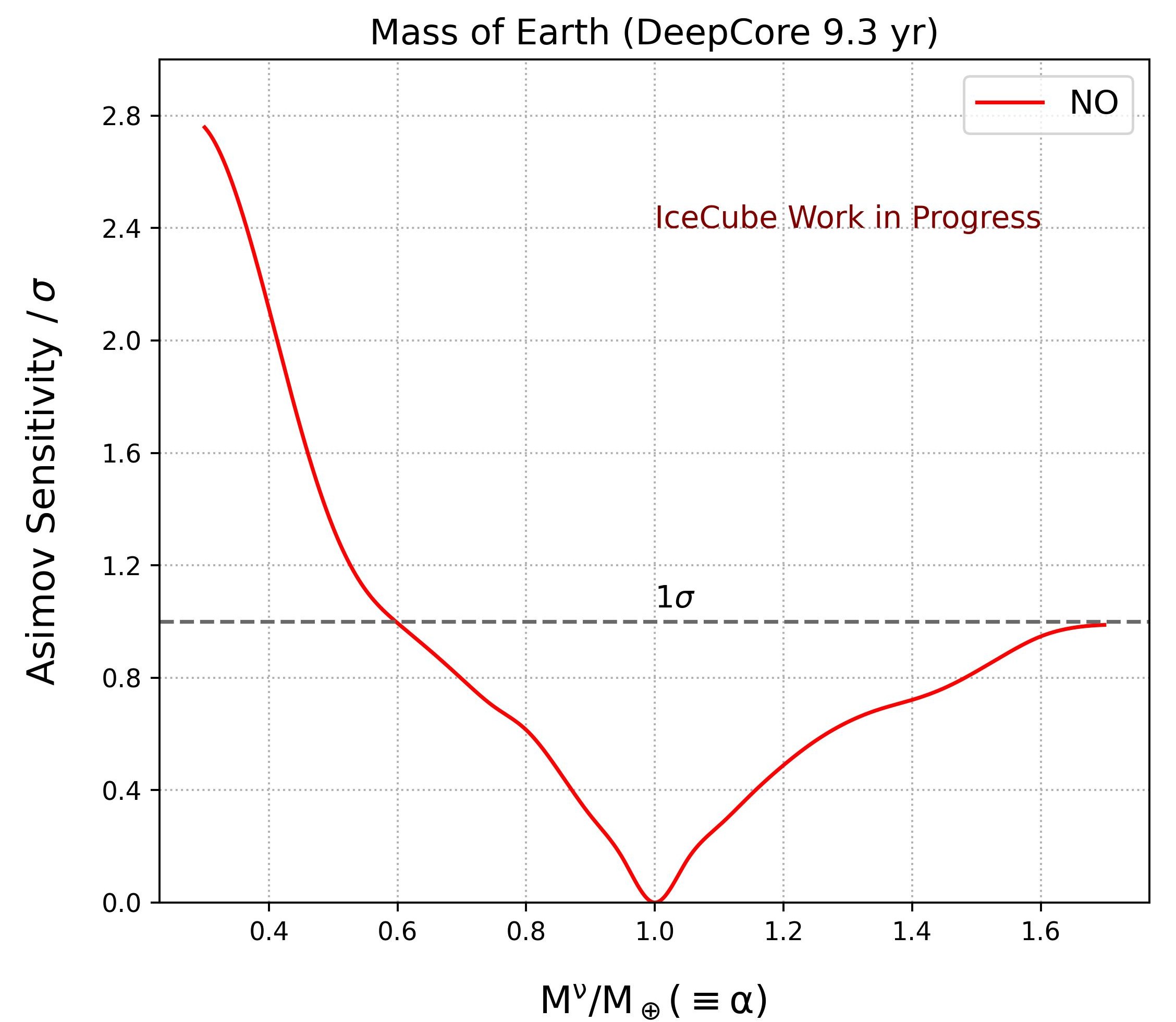}
\includegraphics[width=0.495\linewidth]{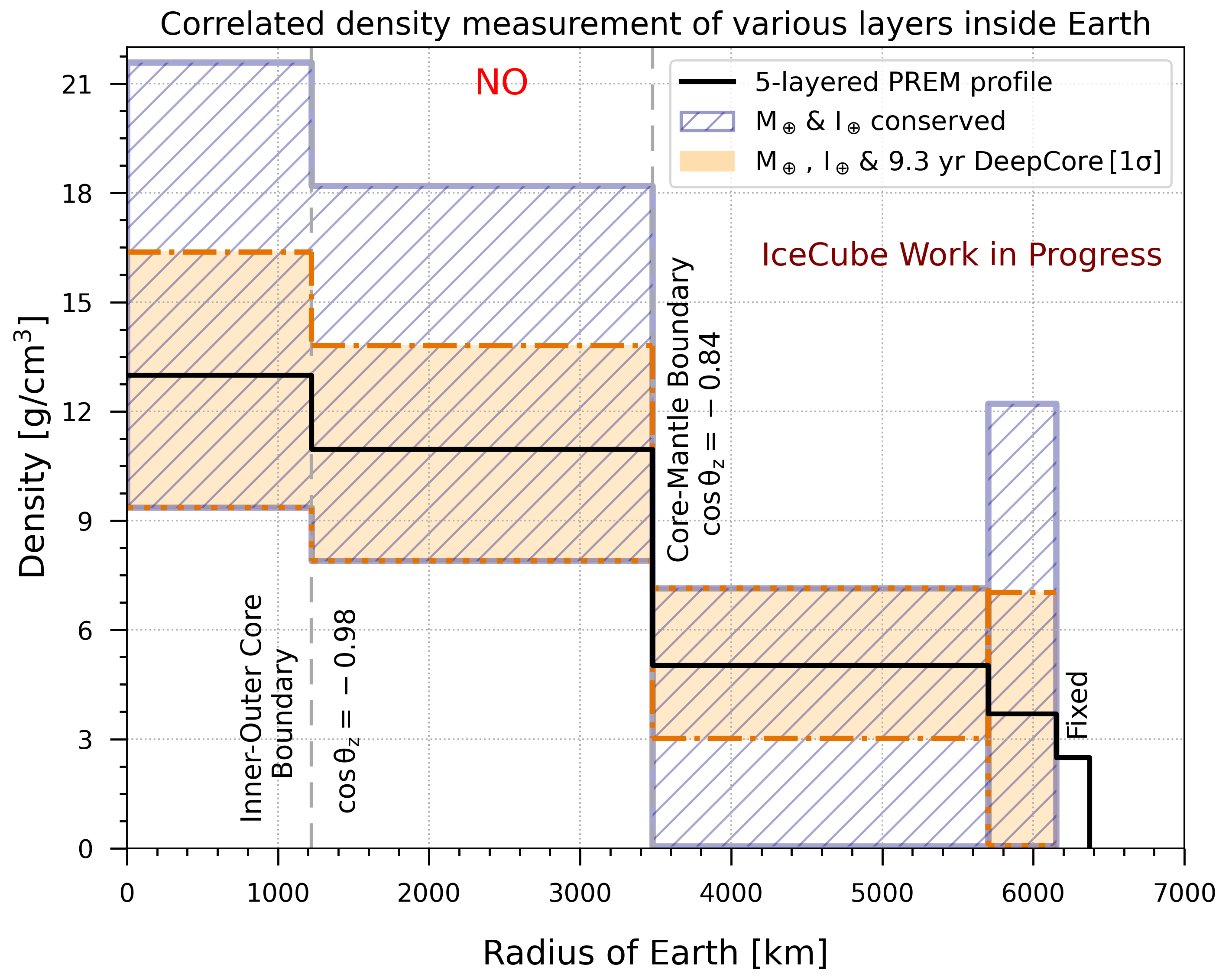}
\caption{Left: Asimov sensitivity for measuring the mass of Earth using atmospheric neutrino oscillations. Here, $M^\nu$ denotes the mass of the Earth as measured by neutrinos, and $M_\oplus$ denotes the gravitationally measured mass of Earth. Right: The orange region shows the 1$\sigma$ Uncertainty around the PREM profile while measuring the correlated densities of various layers inside Earth. This analysis, we combine the simulated neutrino data with the external constraints of mass and moment of inertia of Earth.}
\label{fig:results_analysis3}
\end{figure}

In this section, we present the 1$\sigma$ band for correlated density measurement, obtained from simulated neutrino data combined with external constraints. The right panel of Fig.~\ref{fig:results_analysis3} shows the Asimov sensitivity results of correlated density measurement assuming a 5-layered PREM profile. In the figure, the hatched region shows the band of density allowed by external constraints of mass and moment of inertia. The orange band signifies the 1$\sigma$ band, which we can obtain when we use the simulated data of 9.3 years of the IceCube DeepCore with external constraints of mass and moment of inertia. The dot-dashed orange line indicates the 1$\sigma$ band edges and also signifies the correlation between the density variation of different layers. For example, when the core density increases, the inner mantle density decreases, while the outer mantle density increases simultaneously. This plot indicates that by using neutrino oscillation data in addition to the external constraints of mass and moment of inertia, we can expect an improvement in the parameter space in densities allowed by these constraints.

\section{Conclusions}
\label{sec:conclusion}

We have presented the Asimov sensitivity for observing Earth matter effects in atmospheric neutrino oscillations by rejecting the vacuum oscillation scenario with respect to the matter oscillation scenario using simulated data equivalent to 9.3 years of the IceCube DeepCore detector. We show that, after observing the matter effects in neutrino oscillation data, we are sensitive to the non-uniformity of matter distribution inside Earth by rejecting a uniform density profile with respect to the PREM profile. Further, we present the expected sensitivity to measure the mass of Earth and the correlated densities of different layers of Earth using this simulated neutrino data of the IceCube DeepCore. The next-generation detector, IceCube Upgrade~\cite{IceCube:2023ins} with a lower energy threshold and improved systematics, is expected to enhance the precision of these measurements significantly.

\backmatter

\vspace{0.5cm}
\bmhead{Acknowledgements}
We would like to acknowledge the financial support received from the Department of Atomic Energy (DAE), Government of India. Krishnamoorthi J expresses gratitude to the Science and Engineering Research Board (SERB), Government of India, for the financial assistance provided through the Swarnajayanti Fellowship. Anuj Kumar Upadhyay is thankful to the Department of Science and Technology (DST), Government of India, for the essential financial support offered through the INSPIRE Ph.D. Fellowship.

\vspace{0.5cm}
\bmhead{Data Availability Statement}No data are associated with this manuscript.
\bibliography{sn-bibliography}
\end{document}